\documentclass[preprint,12pt]{elsarticle}
\usepackage{graphicx}
\usepackage{amssymb}
\usepackage{amsmath}
\usepackage{dcolumn}
\usepackage{comment}
\usepackage{color}

\journal{Physica A}

\begin{document}


\title{Hard rigid rods on Husimi lattices}
\author{Lucas R. Rodrigues}
\affiliation{organization={Instituto de Física, Universidade Federal Fluminense},
            addressline={Av. Litorânea s/n}, 
            city={Niterói},
            postcode={24210-346}, 
            state={RJ},
            country={Brazil}}
\author{Tiago J. Oliveira}
\affiliation{organization={Departamento de Física, Universidade Federal de Viçosa},
            city={Viçosa},
            postcode={36570-900}, 
            state={Minas Gerais},
            country={Brazil}}
\author{J\"urgen F. Stilck}
\affiliation{organization={Instituto de Física and National Institute of Science and Technology for Complex Systems, Universidade Federal Fluminense},
            addressline={Av. Litorânea s/n}, 
            city={Niterói},
            postcode={24210-346}, 
            state={RJ},
            country={Brazil}}
\date{\today}

\begin{abstract}
We study the thermodynamic behavior of hard rigid rods of size $k$ (i.e., $k$-mers) on four- and six-coordinated Husimi lattices (HLs), respectively built with squares (square HL) and triangles (triangular HL). In both lattices, dimers ($k=2$) and trimers ($k=3$) only present a isotropic phase, whereas a isotropic-nematic transition is observed for $k \ge 4$. In the square HL, this transition is continuous and occurs at a critical \textit{monomer} activity which displays a nonmonotonic variation with $k$, while the critical \textit{rod} activity and density are always decreasing functions of $k$. The isotropic-nematic transition is discontinuous in the triangular HL, but the $k$-dependence of the coexistence activities and density is analogous to that found for the square case. No transition from the nematic to a high-density disordered phase is found in these HLs. In general, this scenario is very similar to that already observed for rods on the Bethe lattice, though the critical parameters obtained here are in most cases closer to those reported in the literature for the square and triangular lattices. The entropy per site of fully-packed rods is also investigated in detail in the triangular HL, where its value for dimers differs by only 0.7\% from the exact result for the triangular lattice.
\end{abstract}


\maketitle

\section{Introduction}
\label{SecINTRO}

Particle systems with excluded volume interactions only have been studied for much time. Besides a disordered phase, they may also exhibit ordered phases and thus undergo entropy-driven transitions between them. Many different shapes have been considered for the particles, such as hard spheres, squares, triangles, tetraminoes, and so on \cite{cmc05,ps88,fal07,b80,vn99,bsg09,m05}. Some systems composed by binary \cite{binary} and ternary mixtures \cite{ternary} of hard particles have also been considered in the literature. In this paper, we investigate hard rigid rods, which are frequently associated to liquid crystals \cite{vl92}. The study of this particular system started with a seminal paper by Onsager \cite{o49}, where he concluded that, at sufficiently high densities and large aspect ratios, the rods would undergo a discontinuous phase transition between an isotropic and a nematic phase. While Onsager considered the rods in continuous space, Flory \cite{f56} placed them on a lattice, but still allowing their orientations to be continuous. He solved the problem in a mean-field approximation, finding also a transition between isotropic and nematic phases. Zwanzig \cite{z63} studied a system of hard rods in continuum space, but with discrete orientations, finding, as Onsager, a discontinuous isotropic-nematic phase transition.

Here, we are interested in rigid rods placed on lattices, formed by $k$ consecutive monomers aligned in one lattice direction, so that between consecutive monomers of a rod there is a lattice edge. We will call these rods $k$-mers. For the particular case of $k=2$ (dimers) on the square lattice, the entropy of the system in the limit of full coverage has been exactly determined, using both pfaffians \cite{k61,tf61} and transfer matrices \cite{l67,wp21}, being $s_2^{(sq)}=G/\pi = 0.29156 \ldots$ the entropy per site, where $G$ is Catalan's constant. Fully-packed dimers on the triangular lattice have also a long history (see, e.g., Refs. \cite{Nagle66,Phares85,Kenyon00,Fendley02,Wu06}), where $s_2^{(tri)}=0.42859 \ldots$ \cite{Fendley02,Wu06}. The exact determination of the entropy for dimers in this limit is also possible for other two-dimensional lattices \cite{Wu06}, as well as in higher dimensions \cite{dc08}. For larger values of $k$, as far as we known, there are no exact results available for the full lattice entropy of $k$-mers placed on regular lattices, except for the asymptotic $k \rightarrow \infty$ limit, where $s_k = k^{-2}\ln k$ on hypercubic lattices \cite{dr21}. However, rather precise estimates have been obtained using transfer matrix methods for trimers \cite{gdj07} and other $k$-mers with $2 \le k \le 10$ \cite{rsd23} on the square lattice, which agree with the outcomes from numerical simulations \cite{p21}. Semi-analytical solutions of fully-packed rods on generalized Husimi lattices --- built with square lattice clusters of effective lateral size $L$ --- also provide results in fair agreement with those for the square lattice, when extrapolated to the $L \rightarrow \infty$ limit \cite{rso22}. To the best of our knowledgement, there are no analogous results for the triangular lattice reported in the literature.

Turning our attention to the more general case where vacancies may exist between rods placed on the square or triangular lattices, we start mentioning that Monte Carlo simulations demonstrate that no nematic order exists for $k < k_{min} = 7$, whereas a continuous isotropic-nematic transition is found for $k \geq   7$. Its universality class depends on the number of possible orientations of the rods; and may be Ising \cite{mf08,mf08p,fv09} or, more generally, Potts \cite{mf08,mf08p,mf08pp}. The existence of the nematic phase at intermediate densities was also rigorously established for large $k$'s  \cite{ds13}. Interestingly, a second transition  --- from the nematic to a high-density disordered phase --- is found at high rods' densities \cite{gd07}, but its nature is not so well known. Although usual Monte Carlo algorithms using local evaporation and deposition moves are very inefficient in the high density region, a new procedure introduced some time ago (simultaneously updating all sites of a strip using transfer matrix procedures) reduced this difficulty \cite{k12,k13}, indicating a non-Ising transition for $k=7$. However, recent results suggest that this transition is discontinuous for large $k$ \cite{sdr22}.

The thermodynamic behavior of $k$-mers was also studied on the Bethe lattice (BL) and Bethe-like lattices \cite{drs11}. There, a continuous isotropic-nematic transition was found for the four-coordinated lattice ($q=4$) already for $k \ge k_{min}^{(BL)} = 4$; in contrast with $k_{min}=7$ obtained for the square lattice \cite{gd07}. For BLs with coordination $q \ge 6$, the isotropic-nematic transition is discontinuous and appears, again, for $k \ge k_{min}^{(BL)} = 4$; which differs from the continuous transition found (for $k \ge 7$) in the triangular lattice \cite{mf08p}. Furthermore, the transition from the nematic to the high density isotropic phase is missing in the BL for all rod size $k$ and coordination $q$. We recall that the exact (or numerically exact) solution of a given model on the BL corresponds to the result of the so-called mean-field Bethe approximation on a regular lattice with the same coordination number \cite{b82}. So, the differences above lead us to inquire whether the rods' behavior can be better captured by more elaborated mean-field approaches. In order to address this, in this paper, we study the thermodynamic properties of $k$-mers, for general densities, on the four-coordinated Husimi lattice built with squares and on the six-coordinated Husimi lattice built with triangles. Since loops are present in these Husimi cacti, they are expected to provide better mean-field approximations for the models on the square and triangular lattices than the simpler BL solution \cite{h50,t76}. This was indeed the case in the full lattice limit analyzed in Ref. \cite{rso22}, as already mentioned above. As demonstrated in what follows, the scenario found here is not so different to that just described for the BL; even though, in most cases, we obtain critical parameters closer to those estimated in the literature for the regular lattices, provided by techniques such as Monte Carlo simulations or transfer matrix calculations.

The rest of the paper is organized as follows. In Sec. \ref{SecHL} the Husimi lattices are defined and the model is solved on the one built with squares; while the solution on the triangular Husimi cactus is presented in the Appendix. The thermodynamic behavior of rods placed on the square and triangular Husimi lattices are shown and discussed in sections \ref{SecREsquare} and \ref{secREStri}, respectively. Our final discussions may be found in Sec. \ref{SecDI}.

\section{Solution of the model on the Husimi lattice built with squares}
\label{SecHL}

\begin{figure}[ht]
\centering
\includegraphics[width=14.0cm]{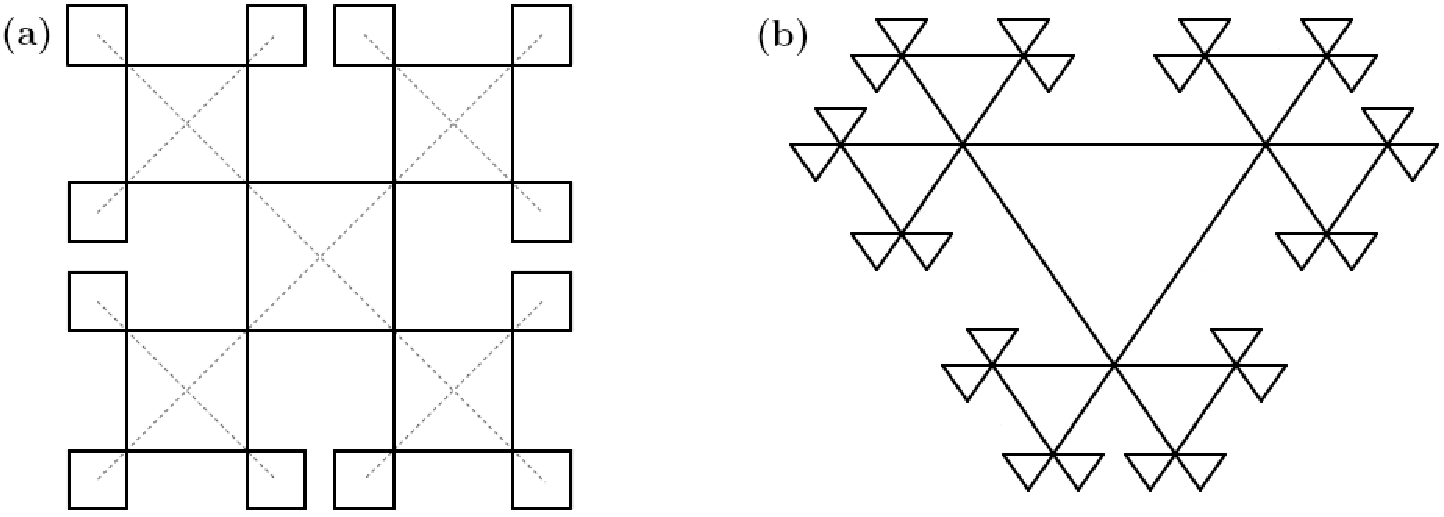}
\caption{Husimi cactuses built with (a) squares and (b) triangles, by starting with a central polygon and adding two generations of them at the surface sites of the previous generations. Following this procedure, such cactuses always have an elementary polygon [square in (a) and triangle in (b)] at their centers. The four-coordinated Cayley tree obtained by connecting the squares' centers in (a) is represented by dashed lines.}
\label{FigHT}
\end{figure}

The Husimi cactus with coordination number $q=4$, as consider here, is a hierarchical structure where successive generations (of shells) of squares are added to a central square, as is shown in Fig \ref{FigHT}(a). A similar cactus, with coordination $q=6$, can be built up by joining generations of triangles to a central triangle, as done in Fig. \ref{FigHT}(b). In contrast to regular lattices, the ratio between the number of surface and bulk sites does not vanish in these cacti even in the thermodynamic limit (i.e., when the number $N$ of generations diverges). However, the core of an infinity Husimi cactus --- i.e., its inner region infinitely far away from the surface --- is expected to form a lattice \footnote{See, e.g., Refs. \cite{b82,ostilli} for clear discussions on this in the context of the Bethe lattice, which is formed by connecting the centers of the polygons forming the Husimi lattices analized here.}, which is known in the literature as a Husimi lattice (HL) \cite{t76}. The solution of a given model in the square HL (triangular HL) illustrated in Fig. \ref{FigHT} provides a mean-field approximation for the model's behavior on the square (triangular) lattice. Notice also that, since surface effects shall become irrelevant in these HLs, the model's properties on them are expected to be shell-independent.

With such considerations in mind, the solution of the rigid rods model on the square HL will be presented here, while the one for the triangular case will be discussed in appendix \ref{secHLtri}. The model we are analyzing is a generalization of the full lattice limit, which was recently considered by some of us on the square HL \cite{rso22}, but now empty sites are allowed and activities $z_x=\exp[\mu_x/(k_B T)]$ and $z_y=\exp[\mu_y/(k_B T)]$ are associated to each \textit{site} occupied by a monomer in a rod aligned in the $x$ and $y$ direction, respectively. The quantities $\mu_x$ and $\mu_y$ are the chemical potentials of a \textit{monomer}, which is part of a $k$-mer, so that we are in the grand-canonical ensemble. Moreover, $k_B$ is Boltzmann's constant and $T$ is the temperature. Although we will restrict ourselves to the thermodynamic properties of the system in the case $z_x=z_y=z$, for latter convenience, the solution will be presented considering different activities. As expected, in all expressions below the full lattice limit results are recovered in the $z \to \infty$ limit.

\begin{figure}[ht]
\centering
\includegraphics[width=7.0cm]{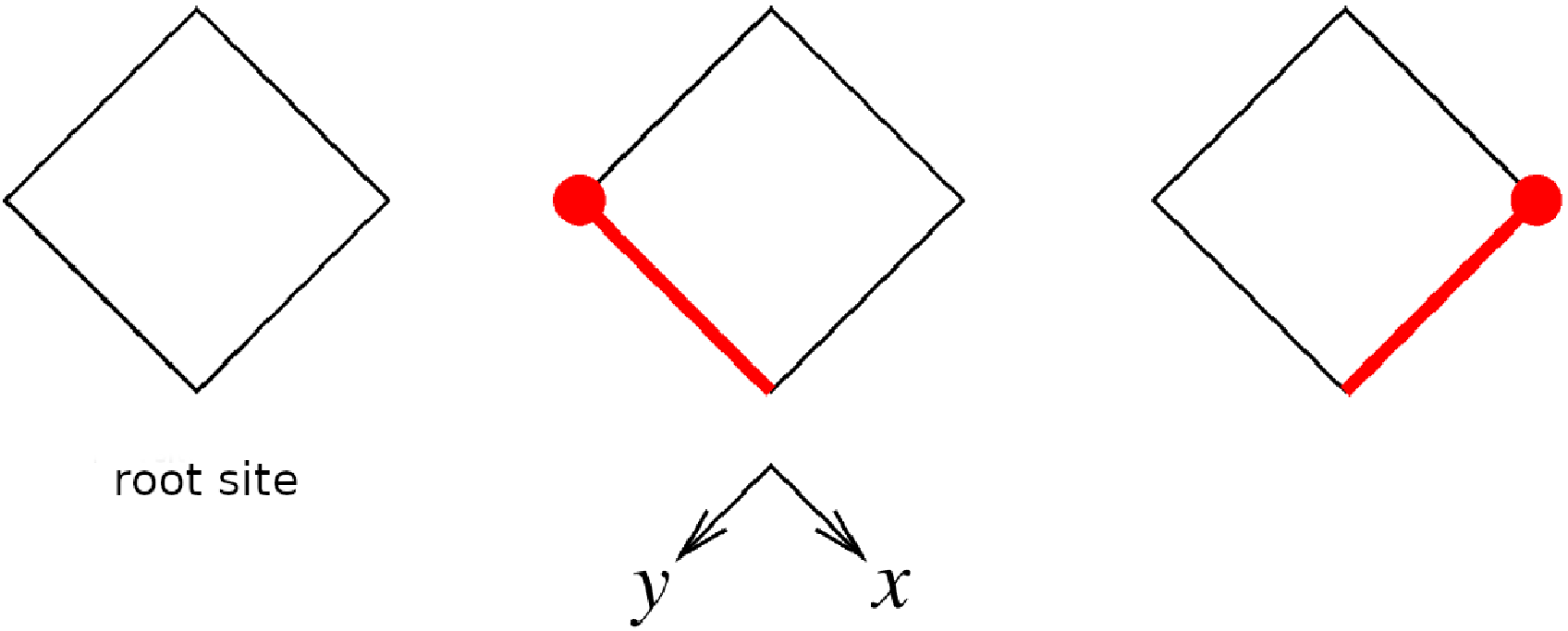}
\caption{Possible root configurations of a sub-cactus. The thick red lines represent bonds of a $k$-mer on the corresponding edge, and the red dots are monomers of rods. Notice that the activity of the monomer which eventually is located at the root site will be considered in the next iteration and, for this reason, it is not represented in the figures. From the left to the right: no rod reaching on the root site from above ($g_0$), a rod in the $x$-direction arriving at the root site ($g_{(i,x)},\,i=1,2,...,k-1$), and a rod in the $y$-direction reaching on the root site ($g_{(i,y)},i=1,2,...,k-1$).}
\label{FigRSTC}
\end{figure}

As usual in solving models on such lattices, we start obtaining recursion relations for the grand-canonical partial partition functions (ppf's) of rooted sub-cactuses for fixed configurations of their root site. In other words, we consider the operation of connecting three sub-cactuses to three corners of a new root square to build up a new sub-cactus with an additional generation of squares. We will label the root configuration of the sub-cactus according to the number of monomers already attached to the rod reaching it from above. If there is no such rod, this label will be equal to zero. If there is a rod reaching it, the label will be the number of monomers already included in the rod with a second component that indicates the direction of the rod, $x$ or $y$. Therefore, there are $2k-1$ ppf's, whose first labels are in the range $[0,...,k-1]$. In Fig. \ref{FigRSTC} the possible sub-cactus root configurations are shown.

\begin{figure}[b]
\centering
\includegraphics[width=7.0cm]{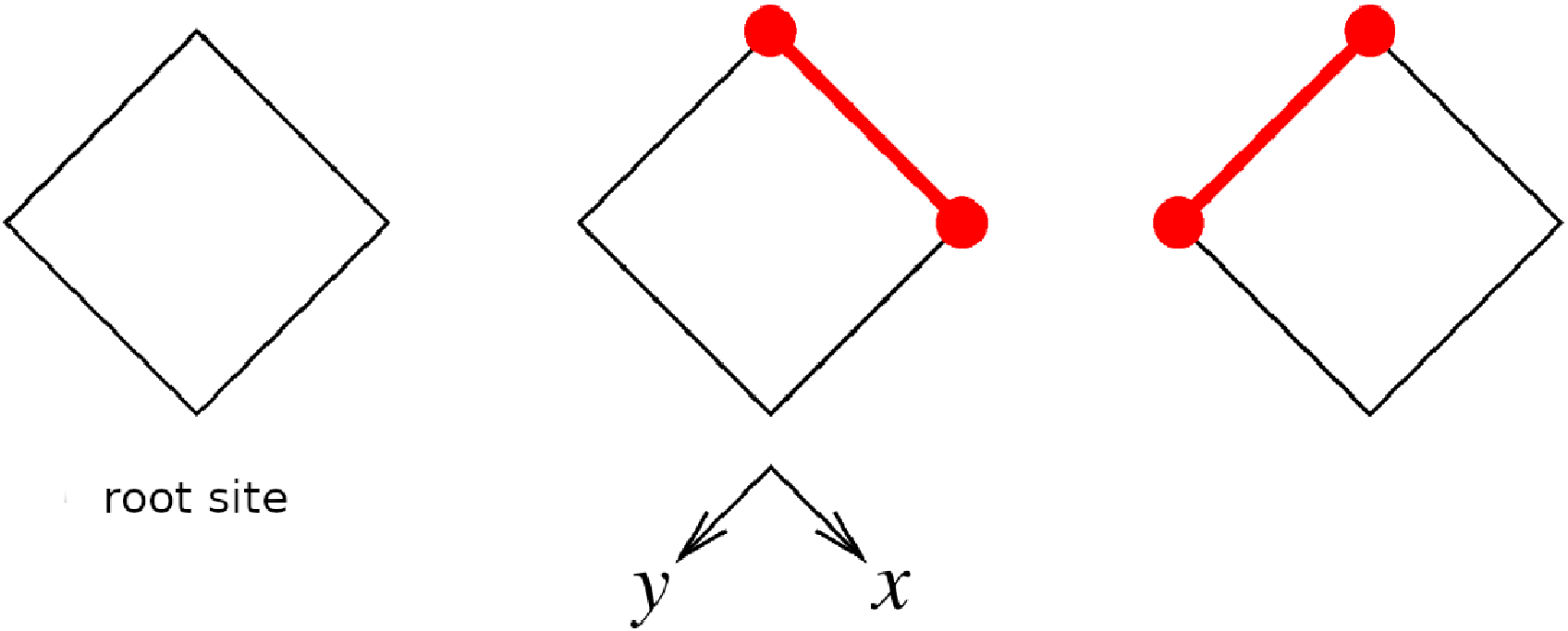}
\caption{Possible configurations of the root square contributing to the recursion relation for $g^\prime_0$. The thick red lines indicate edges of a $k$-mer placed on the new root square, while its monomers are represented by the red dots.}
\label{FigRRST}
\end{figure}

We now proceed obtaining the recursion relations for the ppf's, by considering the operation of attaching three sub-cactuses to a new root square. The possible configurations of the new root square for the case of no rod arriving at the new root site, are illustrated in Fig. \ref{FigRRST}.

Taking into account that the activity of a monomer on the root site of the ppf will be considered in the next iteration, the recursion relation for the partial partition function with no rod reaching on the root site coming from above is:
\begin{eqnarray}
g_0^\prime&=&[g_0+z_x g_{(k-1,x)}+z_yg_{(k-1,y)}]^3+
[g_0+z_xg_{(k-1,x)}+z_y g_{(k-1,y)}]z_x^2\sum_{n=0}^{k-2}g_{(n,x)}g_{(k-2-n,x)}+\nonumber \\
&&+[g_0+z_xg_{(k-1,x)}+z_yg_{(k-1,y)}]z_y^2\sum_{n=0}^{k-2}g_{(n,y)}g_{(k-2-n,y)},
\end{eqnarray}
where it may be helpful to notice that when a bond belonging to a rod occupies one of the two edges in the upper part of the root square, the incoming two parts of this rod should have $n$ and $k-2-n$ monomers, respectively, with $n$ assuming values between $0$ and $k-2$, which explains the sums in the two last terms of the recursion relation. The other recursion relations, for ppf's of sub-cactuses with rods arriving at the root site, may be obtained using similar combinatorial arguments. The results are:
\begin{equation}
    g^\prime_{(i,x)}=z_xg_{(i-1,x)}\left\{[g_0+z_xg_{(k-1,x)}+z_yg_{(k-1,y)}]^2+
    z_x^2\sum_{n=0}^{k-2}g_{(n,x)}g_{(k-2-n,x)}\right\},
\end{equation}
and
\begin{equation}
    g^\prime_{(i,y)}=z_yg_{(i-1,y)}\left\{[g_0+z_xg_{(k-1,x)}+z_yg_{(k-1,y)}]^2+
    z_y^2\sum_{n=0}^{k-2}g_{(n,y)}g_{(k-2-n,y)}\right\},
\end{equation}
where $i=1,2,...,,k-1$. The thermodynamic limit of these ppf's is attained when the number of iterations $N$ diverges, and, as expected, they also diverge in this limit, so it is convenient to define their ratios:
\begin{equation}
    R_{(i,x)}=\frac{g_{(i,x)}}{g_0}, \quad \text{   and   } \quad R_{(i,y)}=\frac{g_{(i,y)}}{g_0}
    \label{rat}
\end{equation}
for $i= 1, 2,...,k-1$. We notice that, if we write the recursion relations for the ppf's as $g^\prime_{(i,x)}=g_0^3f_{(i,x)}$ and $g^\prime_{(i,y)}=g_0^3f_{(i,y)}$, the functions $f_i$ depend only of the activities and the ratios defined above. It is easy to obtain the $2(k-1)$ recursion relations for the ratios. Defining $R_{(i,x)}'=g_{(i,x)}'/{g_0}'$ and $R_{(i,y}'=g_{(i,y)}'/g_0'$, they are:
\begin{equation}
    R^\prime_{(i,x)}=z_xR_{(i-1,x)}\frac{A^2+S_x}{A^3+A(S_x+S_y)},
    \label{rr1}
\end{equation}
and
\begin{equation}
    R^\prime_{(i,y)}=z_yR_{(i-1,y)}\frac{A^2+S_y}{A^3+A(S_x+S_y)},
    \label{rr2}
\end{equation}
where $A=1+z_xR_{(k-1,x)}+z_yR_{(k-1,y)}$, with $i=1,2,...,k-1$ and
\begin{eqnarray}
S_x&=&z_x^2\sum_{n=0}^{k-2}R_{(n,x)}R_{(k-2-n,x)}, \nonumber \\
S_y&=&z_y^2\sum_{n=0}^{k-2}R_{(n,y)}R_{(k-2-n,y)},
\end{eqnarray}
noticing that, from the definition, $R_{(0,x)}=R_{(0,y)}=1$. The recursion relations suggest the following ansatz for the fixed point values of the ratios, attained by them in the thermodynamic limit:
\begin{equation}
  R_{(i,x)}^*=x_1^i \quad \text{    and    }  \quad R_{(i,y)}^*=x_2^i.
  \label{xr}
\end{equation}
Substitution of this ansatz in the fixed point equation $R_{(i,x)}^\prime=R_{(i,x)}$ and $R_{(i,y)}^\prime=R_{(i,y)}$ leads to the following equations:
\begin{subequations}
    \begin{eqnarray}
    x_1[(1+H_{k-1})^3+(k-1)(1+H_{k-1})H_{k-2}]&=&z_x[(1+H_{k-1})^2+(k-1)z_x^2x_1^{k-2}]\\
    x_2[(1+H_{k-1})^3+(k-1)(1+H_{k-1})H_{k-2}]&=&z_y[(1+H_{k-1})^2+(k-1)z_y^2x_2^{k-2}],
\end{eqnarray}
\label{fpe}  
\end{subequations}
where $H_{k-i}=z_x^ix_1^{k-i}+z_y^ix_2^{k-i}$. By either (numerically) solving these equations or iterating the recursion relations (Eqs. \ref{rr1} and \ref{rr2}), for given $z_x=z_y=z$ and $k$, we obtain the fixed points which define the thermodynamic properties of the phases of the model on the square HL.

The grand-canonical partition function $Y_N$ of the system may be obtained considering the operation of connecting four sub-cactuses with $N$ generations to the central square of the Husimi cactus. A convenient alternative is to connect a sub-cactus with $N+1$ generations to another one with $N$ generations, so that they share their root sites. This common root site may be empty or occupied by a monomer of a horizontal or vertical rod, thus:
\begin{equation}
    Y_N=g_0^\prime g_0+\sum_{n=0}^{k-1}(z_xg^\prime_{(n,x)}g_{(k-n-1,x)}+
    z_yg^\prime_{(n,y)}g_{(k-n-1,y)})=g_0^4y,
    \label{pf}
\end{equation}
    where we notice that the sum corresponds to the configurations where a monomer occupies the site shared by the both sub-cactuses, so that $n$ monomers of the rod to which this monomer belongs are on sites of the sub-cactus with $N+1$ generations and $k-n-1$ of its monomers are on sites of the other sub-cactus. The partition function is proportional to $g_0^4$, where it is implicit that this corresponds to a ppf of a sub-cactus with an empty root site and $N$ generations, and the proportionality factor is given by: 
\begin{eqnarray}
    y&=&f_0+\sum_{n=0}^{k-1}(z_xf_{(n,x)}R_{(k-n-1,x)}+z_yf_{(n,y)}
    R_{(k-n-1,y)})=\nonumber\\
    &&f_0\left[1+\sum_{n=0}^{k-1}(z_xR^\prime_{(n,x)}R_{(k-n-1,x)}+
    z_yR^\prime_{(n,y)}R_{(k-n-1,y)})\right],
\end{eqnarray}
where
\begin{equation}
    f_0=A^3+A(S_x+S_y),
    \label{f0}
\end{equation}
The average number of $k$-mers in the horizontal and vertical directions reaching the site of the central square of the cactus where the two sub-cactuses meet are, respectively:
\begin{subequations}
    \begin{eqnarray}
    n_x&=&\frac{z_x\sum_{n=0}^{k-1}R^\prime_{(n,x)}R_{(k-n-1,x)}}
    {1+\sum_{n=0}^{k-1}(z_xR^\prime_{(n,x)}R_{(k-n-1,x)}+
    z_yR^\prime_{(n,y)}R_{(k-n-1,y)})},\\
    n_y&=&\frac{z_y\sum_{n=0}^{k-1}R^\prime_{(n,y)}R_{(k-n-1,y)}}
    {1+\sum_{n=0}^{k-1}(z_xR^\prime_{(n,x)}R_{(k-n-1,x)}+
    z_yR^\prime_{(n,y)}R_{(k-n-1,y)})},
    \end{eqnarray}
    \label{densi}
\end{subequations}
so that the mean density of sites occupied by monomers is
\begin{equation}
    \rho=n_x+n_y,
\end{equation}
and a nematic order parameter may be defined as
\begin{equation}
    \Psi=\frac{|n_x-n_y|}{n_x+n_y}.
\end{equation}

The free energy of the system on the whole Husimi cactus, in the thermodynamic limit, is dominated by the surface. The bulk grand-canonical free energy per site, limited to the central region, discarding the contribution of the surface, is then considered \cite{Gujrati,MinosJurgen,tiagoPol}. Basically, one can assume that the free energy per site of sites located on the surface of the cactus is $\phi_s$ and for sites in the bulk it is equal to $\phi_b$. Then, one applies this assumption to two cactuses, with $N$ and $N+1$ generations respectively, which allows the determination of $\phi_b$ (as well as $\phi_s$). The result for the bulk free energy per site is:
\begin{equation}
    \phi_b=-\frac{k_BT}{4}\ln\left[\frac{Y_{N+1}}{Y_N^3}\right].
\end{equation}
We remark that the same result can be obtained, in the $N \rightarrow\infty$ limit, if one assumes that $\phi$ is a function of the cactus shell, being $\phi = \phi_b$ at the center, as above \cite{tiago10}. It may then be shown that
\begin{equation}
   \varphi_b \equiv \frac{\phi_b}{k_B T}=-\ln\left[\frac{f_0}{y^{1/2}}\right].
    \label{fe}
\end{equation}
In the thermodynamic limit ($N \to \infty$) the ratios should be replaced by their fixed point values $R_{(i,x)}^*=x_1^i$ and $R_{(i,y)}^*=x_2^i$, where $x_1$ and $x_2$ are the solutions of the fixed point equations (\ref{fpe}).

In the calculations above we defined the Husimi cactus so that the center of an elementary square is located at its center. We could alternatively define it in such a way that a site of the cactus is located at the center. In this case, although we would expect different results for a finite cactus, the average values at the central region of both cactuses will be equal in the thermodynamic limit. The partition function in this case is obtained connecting two sub-cactuses with the same number of generations to the central site, giving
\begin{equation}
    Y_{a,N}=g_0^2+\sum_{n=0}^{k-1}[z_xg_{(n,x)}g_{(k-n-1,x)}+
    z_yg_{(n,y)}g_{(k-n-1,y)}]=g_0^2y_a,
\end{equation} 
where now
\begin{equation}
    y_a=1+\sum_{n=0}^{k-1}[z_xR_{(n,x)}R_{(k-n-1,x)}+z_yR_{(n,y)}R_{(k-n-1,y)}].
\end{equation}
The average number of $k$-mers in the horizontal and vertical directions reaching the central site in the thermodynamic limit may be expressed in terms of the parameters $x_1$ and $x_2$ (Eq. \ref{xr}) as
\begin{subequations}
\begin{eqnarray}
n_x=&=&\frac{z_xkx_1^{k-1}}{1+k(z_xx_1^{k-1}+z_yx_2^{k-1})}, \\
n_y=&=&\frac{z_ykx_2^{k-1}}{1+k(z_xx_1^{k-1}+z_yx_2^{k-1})}.
\end{eqnarray}
\label{densix}
\end{subequations}
and a procedure similar to the one used above for the square-centered cactus leads to the bulk free energy per site:
\begin{equation}
   \varphi_b \equiv \frac{\phi_b}{k_B T} =-\frac{1}{2}\ln\left[\frac{f_0}{y_a}\right],
    \label{fea}
\end{equation}
where $f_0$ is given in Eq. \ref{f0}.

One point which is worth looking at is whether the free energies per site [in Eqs. (\ref{fe}) and (\ref{fea})] are consistent with the ones that would be obtained integrating the densities with respect to the chemical potential. So, considering that
\begin{equation}
    n_x=-k_BT\left(\frac{\partial \phi_b}{\partial \mu_x}\right)_{\mu_y,T}=
    -z_x\left(\frac{\partial \phi_b}{\partial z_x}\right)_{z_y},
\end{equation}
and
\begin{equation}
    n_y=-z_y\left(\frac{\partial \phi_b}{\partial z_y}\right)_{z_x},
\end{equation}
we have calculated the densities of monomers in $k$-mers in both lattice directions directly from the free energy in Eq. (\ref{fea}) and compared the results with those provided by Eqs. (\ref{densix}) --- for several values of the activity per monomer $z$ and of the number of monomers in rods $k$, both in the isotropic and in the nematic phase ---, finding that they coincide within the numerical precision we are working.

The entropy per site may be obtained considering that $\phi_b=u_b-T s - \rho\mu$, so that
\begin{equation}
    \frac{s}{k_B}=-\varphi_b-\rho\ln z,
\end{equation}
for $u_b=0$, since the system is athermal.

Finally, to study the stability of the fixed points, we obtained the $(2k-2) \times (2k-2)$ Jacobian matrix of the recursion relations of the ratios of ppf's, whose elements are:
\begin{equation}
    J_{i,j}=\left(\frac{\partial R^\prime_i}{R_j}\right).
\end{equation}
A particular fixed point will be stable if the modulus of the leading eigenvalue of the Jacobian is smaller than one.

\section{Results for rods on the square HL}
\label{SecREsquare}

To obtain the thermodynamic properties of the model on the Husimi lattice, we start, for fixed values of the monomer activity $z$ and number of monomers in the rod $k$, calculating the fixed points of the recursion relations in Eqs. \ref{rr1} and \ref{rr2}. This may be done by iterating such recursion relations, but it is simpler to solve the fixed point equations \ref{fpe} and the ratios may then be obtained through expressions \ref{xr}. For $k=2,3$, we find that only the isotropic fixed point exists ($x_1=x_2=x)$, for any value of the activity $z$, so that the nematic order parameter vanishes identically. For $k \ge 4$ an isotropic fixed point is stable for small activities, while for higher activities a nematic fixed point is stable. The stability limits of both fixed points are coincident [at a critical activity $z_c(k)$], indicating that the transition between the isotropic and the nematic phase is continuous.  

\begin{figure}[t]
\centering
\includegraphics[width=7.0cm]{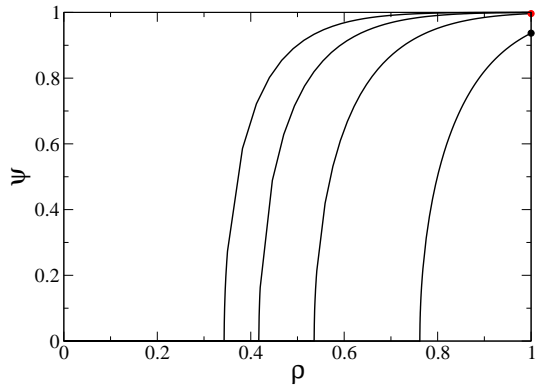}
\caption{Nematic order parameter as a function of the density of sites occupied by monomers, for $k$-mers placed on the square HL. From right to left, results for $k=4,5,6,$ and $7$.}
\label{psi}
\end{figure}

This is indeed confirmed in Fig. \ref{psi}, where the order parameter is shown as a function of the density of sites occupied by the rods on the lattice, for different values of $k$. As $k$ grows, $\Psi$ approaches unity in the limit of full occupancy (corresponding to $z \rightarrow \infty$ and, thus, $\rho=1$), but it does not reach it, as already demonstrated in Ref. \cite{rso22}. Therefore, in such a limit the stable phase is nematic for all $k \ge 4$ and no nematic-to-isotropic transition exists at high densities, similarly to the BL solution \cite{drs11}.

The values of the critical activities, densities, and entropies for $4 \le k \le 15$, are summarized in Table \ref{TabHLsq}. We notice that the critical densities and entropies are monotonically decreasing functions of $k$, but the critical activities show a minimum at $k=10$, followed by a slow increase.  We may compare the critical parameters  with the ones found for the model on the four coordinated Bethe lattice in \cite{drs11}, which are:
\begin{equation}
    z_{c}^{(BL)}=\frac{(k-1)^{\frac{2k-2}{k}}}{k(k-3)},
\end{equation}
\begin{equation}
    \rho_{c}^{(BL)}=\frac{2}{k-1},
\end{equation}
and
\begin{eqnarray}
    s_{c}^{(BL)}&=&2\left(1-\frac{1}{k}\right)\ln\left(1-\frac{1}{k}\right)-
    \left(1-\frac{2}{k-1}\right)\ln\left(1-\frac{2}{k-1}\right)- \nonumber \\
    && 2\left(\frac{1}{k(k-1)}\right)\ln\left(\frac{1}{k(k-1)}\right).
\end{eqnarray}
We recall that the Bethe and Husimi lattice solutions of a model are mean-field approximations, and the second one is of higher order than the first. It is thus interesting to compare the results of both calculations and eventual other available estimates, as the ones provided by simulations. We  found that, as expected, the absolute values of the relative differences between the values of these parameters on the two lattices are decreasing functions of $k$, starting with values of the order of 20\% for $k=4$ and ending with values around 0.5\% or lower.

\begin{table}[t]
\caption{Critical activity ($z_c$), density ($\rho_c$) and entropy ($s_c$) for rods with different number of monomers $k$ placed on the square HL.}
\label{TabHLsq}
\centering
\begin{tabular}{cccc}
$k$   &     $z_c$    &    $\rho_c$   &    $s_c$   \\
\hline
4     &    1.675089    &    0.761905     &    0.320403        \\ 
5     &    0.964988    &    0.535714     &    0.293498        \\ 
6     &    0.827138    &    0.417391     &    0.233700        \\ 
7     &    0.777397    &    0.343137     &    0.186956        \\ 
8     &    0.756853    &    0.291793     &    0.152346        \\ 
9     &    0.748910    &    0.254032     &    0.126464        \\ 
10    &    0.747170    &    0.225035     &    0.106714        \\ 
11    &    0.748709    &    0.202040     &    0.091330        \\
12    &    0.752040    &    0.183346     &    0.079117        \\
13    &    0.756352    &    0.167840     &    0.069258        \\
14    &    0.761179    &    0.154767     &    0.061181        \\
15    &    0.766244    &    0.143594     &    0.054477
\end{tabular}
\end{table}

\begin{figure}[t]
\centering
\includegraphics[width=7.0cm]{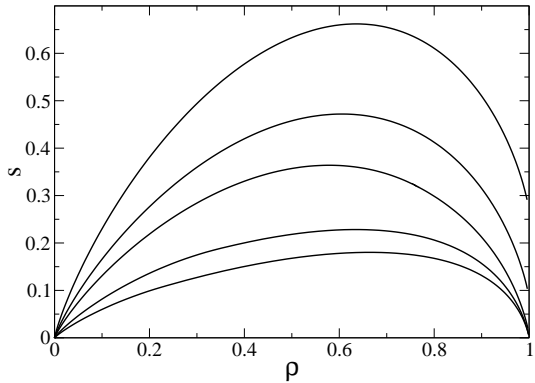}
\caption{Entropy per lattice site (in units of $k_B$) as a function of the density of sites occupied by monomers, for $k$-mers placed on the square HL. Downwards: results for $k=2, 3, 4,$ and $7$.}
\label{ent}
\end{figure}

The entropy per lattice site, also as a function of the density of sites occupied by monomers, is depicted in Fig. \ref{ent}. It obviously starts at zero for $\rho=0$, reaches a maximum at an intermediate density, which grows monotonically with $k$, and ends at the values already obtained in Ref. \cite{rso22} in the full lattice limit. 

As expected, when we compare the results found on the Husimi lattice with the behavior of the model on the Bethe lattice \cite{drs11}, in general, the results on the Husimi lattice are somewhat closer to the estimates provided by numerical simulations of the system on the square lattice. For instance, as discussed in the Introduction, on the square lattice the nematic order parameter vanishes in the full lattice limit, while both in the Bethe and Husimi lattices it does not vanish. However, as demonstrated in \cite{rso22}, $\Psi$ is always smaller on the Husimi lattice than on the Bethe lattice; see table III of Ref. \cite{rso22}. It is also noteworthy in such a table, as it was seen in the comparisons between critical parameters on the Bethe and Husimi lattices above, that the difference between the order parameters on the two lattices becomes smaller as $k$ increases, since the ratio between the size of the rods and the size of the elementary squares of the lattice grows. This effect is observed in all comparisons between both lattices, being clearly visible, e.g., in the full lattice entropies presented in table III of Ref. \cite{rso22}.

\begin{figure}[t]
\centering
\includegraphics[width=7.0cm]{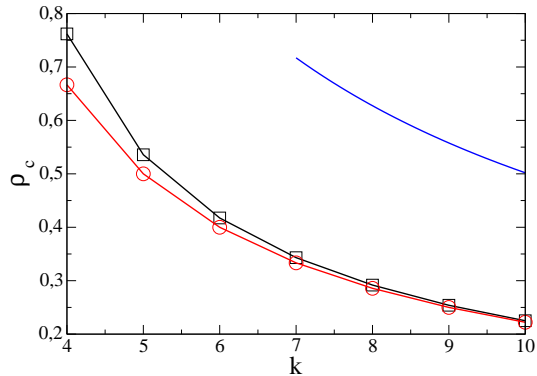}
\caption{Critical densities on Bethe and square HL, as functions of $k$. Circles correspond to the four coordinated Bethe lattice results \cite{drs11}, squares to Husimi lattice results and the blue line are simulational estimates for the model on the square lattice \cite{mf08,mf08p}}. 
\label{rhoc}
\end{figure}

The values of the critical density on the HL as a function of the size of the rod $k$ are compared with those for the BL and for the square lattice in Fig. \ref{rhoc}. The latter ones were obtained by Matoz-Fernandez \textit{et al.} \cite{mf08,mf08p}, according to which $\rho_c(k) \sim k^{-1}$ and, moreover, $\rho_c(10) \approx 0.502$ for rigid rods on the square lattice, such that $\rho_c(k) \simeq 5.02/k$. We notice that, for $k=7$ this expression gives $\rho_c(7) \approx 0.717$, which is not so different from the estimate reported by Kundu \textit{et al.} \cite{k13}: $\rho_c(7) \approx 0.745$. Recalling that results on these hierarchical lattices are mean-field approximations of the model on the square lattice, we expect them to underestimate the critical density. This is indeed confirmed in Fig. \ref{rhoc}, where we may see that the critical densities on the square HL are larger than those found on the Bethe lattice. Of course, the mean-field results are still much lower than the available estimates for the square lattice ones.

\section{Results for rods on the triangular HL}
\label{secREStri}

\subsection{Isotropic-nematic transition}

\begin{figure*}
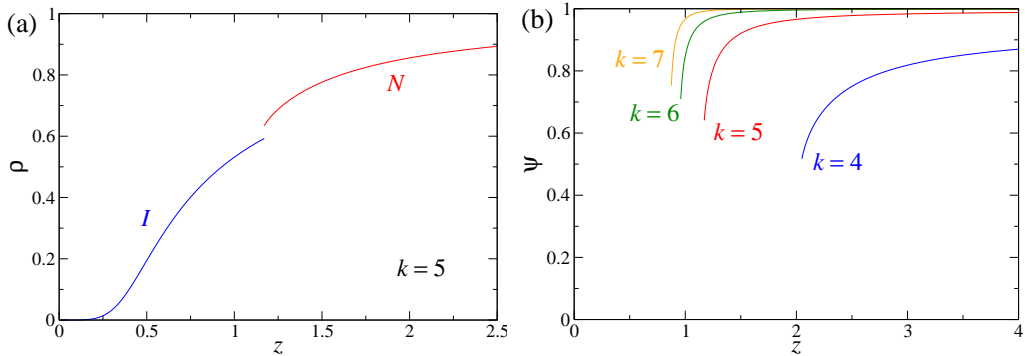

\centering
\includegraphics[width=6.70cm]{FigHLtriRho.eps}
\includegraphics[width=6.70cm]{FigHLtriPsi.eps}
\caption{Results for rigid rods placed on the triangular HL. (a) Monomer density $\rho$ versus activity $z$, for rods with $k=5$, in the isotropic ($I$) and nematic ($N$) phases. (b) Nematic order parameter $\Psi$ as a function of $z$, for the indicated $k$'s. Similar results are found for all $k \geq 4$.}
\label{FigresHLtri}
\end{figure*}

Similarly to the results above for the square HL (and also to those for the BL \cite{drs11}), the $k$-mers never present nematic order in the triangular HL for $k=2$ and $3$, whereas an isotropic-nematic transition is found for $k \geq 4$. Such a transition is always discontinuous in the triangular HL case, in consonance with the BL approach for coordination $q=6$ \cite{drs11}. This is shown in Fig. \ref{FigresHLtri}, where one sees that both the monomer density $\rho$ and the order parameter $\Psi$ present a discontinuous behavior as $z$ increases.

\begin{figure*}
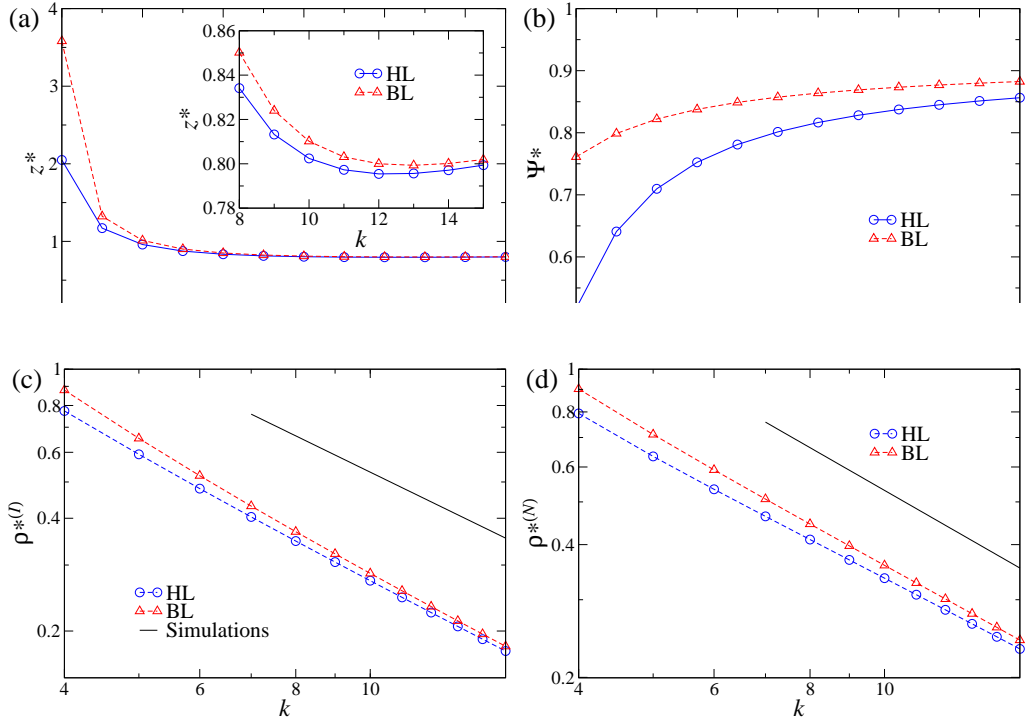

\centering
\includegraphics[width=6.70cm]{FigHLtriZcoex.eps}
\includegraphics[width=6.70cm]{FigHLtriPsicoex.eps}
\includegraphics[width=6.70cm]{FigHLtriRhoIcoex.eps}
\includegraphics[width=6.70cm]{FigHLtriRhoNcoex.eps}
\caption{Comparison between results for $k$-mers placed on the triangular HL and on the BL, at the coexistence points: (a) activity $z^*$, (b) monomer density of the nematic phase $\rho^{*(N)}$ and (c) nematic order parameter $\Psi^*$ versus $k$. The insertion in panel (a) highlights the behavior of $z^*$ for large $k$.}
\label{FigresHLtri2}
\end{figure*}

The coexistence points $z^*$ --- estimated from the condition $\phi_b^{(I)}=\phi_b^{(N)}$, for a given $k$ --- are summarized in Table \ref{TabHLtri} and displayed also in Fig. \ref{FigresHLtri2}(a). As was found for the critical activities on the square HL, $z^*$ displays a non-monotonic variation with $k$: it decreases fast for small rods; then, presents a minimum at $k=12$; after which it passes to grow very slowly. On the other hand, the monomer densities at the coexistence, $\rho^{*(I)}$ and $\rho^{*(N)}$, are always decreasing functions of $k$ (at least for $k \leq 20$, which was the largest case analyzed here), as seen in Tab. \ref{TabHLtri} and Figs. \ref{FigresHLtri2}(c) and \ref{FigresHLtri2}(d). The gap between these densities $\Delta \rho^* = \rho^{*(N)} - \rho^{*(I)}$ is small, being $\Delta \rho^* \lesssim 0.06$, though the relative difference $\Delta \rho^*/\rho^{*(N)}$ increases monotonously with $k$, since $\rho^{*(N)}$ decreases faster than $\Delta \rho^*$.

As may be observed in Fig. \ref{FigresHLtri}(b), in Table \ref{TabHLtri} and in Fig. \ref{FigresHLtri2}(b), the jump in the order parameter at coexistence, $\Psi^*$, increases with $k$, suggesting that the discontinuous character of the transition gets stronger for larger $k$'s. In contrast, however, the coexistence region decreases with $k$. Indeed, the activities, $z_s$, at the spinodals (estimated from the largest eigenvalue of Jacobian $|\Lambda|=1$) of the isotropic and nematic phases are also displayed in Tab. \ref{TabHLtri}, where one observes that $\Delta z_s = z_s^{(I)} - z_s^{(N)}$ decreases (very slowly asymptotically) with the rods' size.

Noteworthy, the scenario above is very similar to that found for rods on the BL with $q=6$ \cite{drs11}. It is important to remark, however, that the coexistence points were estimated in Ref. \cite{drs11} by writing down the entropy [Eq. (32) there] as a function of $\rho$ and $\Psi$, and then defining the ``critical density" $\rho_c$ as the value for which $s(\Psi,\rho=\rho_c)$ has two peaks of equal height \footnote{When comparing the results here with those in Ref. \cite{drs11}, it is important to bear in mind that the activity there (let us call it $\tilde{z}$) is for rod, while here $z$ is for a single monomer, such that $z=\tilde{z}^{1/k}$.}. We have revisited such calculations for the six-coordinate BL here, but determining the coexistence points through the free energy --- i.e., by requiring that $\phi_b^{(I)}=\phi_b^{(N)}$ ---, which gives results slightly different from those reported in \cite{drs11}.

\begin{table}[t]
\caption{Coexistence activity ($z^{*}$), densities ($\rho^{*(I)}$ and $\rho^{*(N)}$) and order parameter ($\Psi^*$), and stability limits ($z_s^{(I)}$ and $z_s^{(N)}$) for rods of different sizes $k$ placed on the triangular HL.}
\label{TabHLtri}
\centering
\begin{tabular}{ccccccc}
$k$   &     $z^*$    & $\rho^{*(I)}$ &  $\rho^{*(N)}$  &    $\Psi^*$    &    $z_s^{(I)}$   &  $z_s^{(N)}$       \\
\hline
4     &    2.049172  &    0.773600   &    0.793726     &    0.517008    &    2.347627      &    2.015605        \\ 
5     &    1.170803  &    0.592492   &    0.634234     &    0.640842    &    1.285300      &    1.157497        \\ 
6     &    0.959370  &    0.479949   &    0.534063     &    0.709700    &    1.035761      &    0.950479        \\ 
7     &    0.874695  &    0.403266   &    0.463657     &    0.752365    &    0.934285      &    0.867787        \\ 
8     &    0.834113  &    0.347676   &    0.410778     &    0.780991    &    0.883966      &    0.828363        \\ 
9     &    0.813224  &    0.305538   &    0.369300     &    0.801365    &    0.856576      &    0.808250        \\ 
10    &    0.802429  &    0.272499   &    0.335746     &    0.816533    &    0.841056      &    0.798020        \\ 
11    &    0.797270  &    0.245902   &    0.307967     &    0.828226    &    0.832262	   &    0.793294        \\ 
12    &    0.795437  &    0.224031   &    0.284549     &    0.837496    &    0.827522      &    0.791806        \\ 
13    &    0.795638  &    0.205731   &    0.264514     &    0.845015    &    0.825328      &    0.792290        \\ 
14    &    0.797109  &    0.190193   &    0.247165     &    0.851228    &    0.824783      &    0.793999        \\ 
15    &    0.799378  &    0.176836   &    0.231985     &    0.856445    &    0.825325      &    0.796471        \\ 
\end{tabular}
\end{table}

Figures \ref{FigresHLtri2}(a)-(d) present the outcomes from such BL solution, comparing them with those for the triangular HL. In general, all quantities have larger values on the BL, but the same qualitative behavior. For instance, $z^*$ has also a non-monotonic variation with $k$ (with a minimum at $k=13$) in the BL case. The same non-monotonic behavior is observed in the spinodals of both the triangular HL (see Tab. \ref{TabHLtri}) and the BL. In fact, for the BL the spinodal of the isotropic phase can be exactly determined, being \cite{drs11}
\begin{equation}
z_s^{(I,BL)} = \frac{\left(1-\frac{1}{k}\right)^{\frac{k-1}{k}}}{\left(1-\frac{3}{k-1}\right)[k(k-1)]^{\frac{1}{z}}},
\end{equation}
which displays a minimum at $k = 16$, considering that $k \in \mathbb{Z}$. It is interesting that $z_s^{(I,BL)} \rightarrow \infty$ for $k=4$, meaning that the isotropic phase is always stable (or metastable) in the BL. In the triangular HL, on the other hand, $z_s^{(I)}$ is finite (and small) for all $k \ge 4$ (see Tab. \ref{TabHLtri}). Actually, for all $k \ge 4$, we find that the size of the coexistence region $\Delta z_s = z_s^{(I)} - z_s^{(N)}$ is larger in the BL case.

Moreover, although the transitions are discontinuous in both the BL and triangular HL solutions, the nematic order parameters at coexistence are much smaller in the HL case, especially for the smaller $k$'s. These results demonstrate that, by improving the approximation, one obtains a behavior closer to a continuous transition, as expected for the triangular lattice \cite{mf08,mf08p}. Curiously, however, the coexistence densities do not follow this trend. In fact, from Monte Carlo simulations of rigid rods on the triangular lattice and some theoretical approaches, Matoz-Fernandez \textit{et al.} \cite{mf08,mf08p} found that the continuous isotropic-nematic transitions occur at critical densities given by $\rho_c(k) \simeq 5.31/k$, for $k \ge 7$. As it may be observed in Figs. \ref{FigresHLtri2}(c) and \ref{FigresHLtri2}(d), these critical densities are much larger than the coexistence ones found here. Moreover, both $\rho^{*(I)}(k)$ and $\rho^{*(N)}(k)$, are more distant from $\rho_c(k)$ in the HL case.

\subsection{The full-packing limit}

Now, we will discuss the case where all sites of the triangular HL are occupied by rods, corresponding to the $z \rightarrow \infty$ limit (and $\rho=1$) in the solution presented in the appendix.

The entropies of the isotropic and nematic phases are presented in Table \ref{TabHLtriCheia}, for $2 \leq k \leq 7$. In agreement with the results from the previous subsection, only the isotropic phase is stable in the full lattice case for dimers and trimers. For larger $k$'s, however, the entropy of the isotropic phase is negative, indicating that this phase is unstable, while that for the nematic phase it is positive. Therefore, for $k \geq 4$, the nematic phase is the single one stable for large $z$ and this persists until the fully occupancy limit. Hence, no nematic-isotropic transition close to full-packing --- as observed for rigid rods on the square lattice \cite{k12} --- occurs in the triangular HL. The nematic order parameters are also shown in Tab. \ref{TabHLtriCheia}, which are close, but smaller than 1. In addition, in agreement with previous results for the Bethe lattice (BL) \cite{drs11} and the square HL \cite{rso22}, they get close to $\Psi = 1$ as $k$ increases.

For comparison, Table \ref{TabHLtriCheia} presents also the entropies and order parameters for rigid rods placed on the BL with coordination $q=6$, in the full lattice limit, as obtained in Ref. \cite{drs11}. As in the HL case, the stable phase is isotropic for $k \leq 3$ and nematic for $k \geq 4$. Interestingly, the entropies of these phases are very similar in the BL and HL approximations. For instance, they differ by less than 2\% in the case of dimers, for which they are very close also to the exact value on the triangular lattice: $s_2^{(tri)}=0.428 594 537\ldots$ (see, e.g., Ref. \cite{Wu06} and references therein). Remarkably, the difference between the HL and the exact result for dimers is only 0.7\%. This is certainly a consequence of the fact that the dimer-dimer correlations on the triangular lattice are short ranged \cite{Fendley02}, so that the weakening of correlations caused by the hierarchical structure of the HL has a small effect on the properties of this model. Also, this suggests that generalized triangular HLs --- built, e.g., with hexagonal triangular lattice cells, in the same spirit of the generalized square HLs considered in Refs. \cite{ro21,rso22} --- may give the exact solution of the dimer problem for finite (and perhaps small) cells.

\begin{table}[t]
\caption{Entropy $s$ of fully-packed rods placed on the Bethe lattice (BL) with coordination $q=6$ and on the triangular HL, for the isotropic ($I$) and nematic ($N$) phases, and several $k$'s. The values of the nematic order parameter $\Psi$ are also shown into parentheses.}
\label{TabHLtriCheia}
\centering
\begin{tabular}{ccccccc}
$k$             &     2        &      3        &       4            &       5           &       6         &       7          \\
\hline
$s_{BL}^{(I)}$  &    0.44007   &    0.14600    &    -2.6058E-2      &   -0.14073        &    -0.22335     &   -0.28608       \\ 
$s_{BL}^{(N)}$  &    --        &    --         &     9.3628E-3      &    6.5788E-4      &     4.3011E-5   &    2.4293E-6     \\
$(\Psi)$        &              &               &     (0.93030)      &    (0.99492)      &     (0.99961)   &    (0.99997)    \\ 
\hline
$s_{HL}^{(I)}$  &    0.43152   &    0.11896    &    -7.1920E-2      &   -0.20349        &    -0.30088     &   -0.37646        \\ 
$s_{HL}^{(N)}$  &    --        &    --         &     8.6686E-3      &    6.5078E-4      &     4.2953E-5   &    2.4289E-6      \\
$(\Psi)$        &              &               &     (0.94162)      &    (0.99503)      &     (0.99961)   &    (0.99997)      \\ 
\end{tabular}
\end{table}

As demonstrated in Table III of Ref. \cite{rso22}, which shows the same results as presented in Table \ref{TabHLtriCheia} here, but for the BL with $q=4$ and for the square HL --- the rods' entropies in these four-coordinated lattices are smaller than those in the six-coordinated cases analyzed here, as expected. Moreover, while the entropy for a given $k$ converges from below in $q=4$ case (see Tabs. III and V, and Fig. 2 in Ref. \cite{rso22}), for the lattices considered here (with $q=6$) it converges from above to the asymptotic (triangular lattice) behavior. Namely, the number of possible $k$-mers' configurations decreases as one goes from the BL ($q=6$) to the triangular HL and, then, to the triangular lattice; while the opposite behavior is observed for the lattices with $q=4$.

\section{Discussion}
\label{SecDI}

We have studied the model of $k$-mers, rigid rods composed by $k$ monomers placed on lattice sites with $k-1$ edges of the lattice between them, all aligned in the same direction, with excluded volume interactions only; that is, only configurations of the rods where each lattice site is occupied by at most one monomer are accepted. We obtained the solution of this model in the grand-canonical ensemble in the central region of two hierarchical lattices built with polygons (Husimi cactuses): the cactus built with squares, so that two squares meet at each lattice site (ramification of squares equal to one and coordination number $q=4$); and the one built with triangles, with three of them meeting at each site (ramification of triangles equal to two and $q=6$). The exact solution in the core of these cactuses corresponds to a mean-field approximation of the model on the square and triangular lattice, respectively. Due to the existence of loops in the Husimi lattices, these calculations may be considered to be higher order mean-field approximations to the problem as compared to the Bethe lattice solution, which was obtained in Ref. \cite{drs11}.

On the Husimi lattice built with squares, the rods display a continuous isotropic-nematic transition, as it is also the case on the four-coordinated Bethe lattice and on numerical simulations on the square lattice. In contrast, a discontinuous transition was found for the model on the Husimi lattice built with triangles, as happens also on the six-coordinated Bethe lattice, while numerical simulations on the triangular lattice suggest a continuous transition in the class of the 3-state Potts model. Therefore, in this case, the Husimi lattice solution also does not lead to the order of the transition obtained using numerical simulations. It is worthy remarking, however, that mean-field approaches to the 3-state Potts model are known to display a discontinuous transition \cite{w82}, which suggests that the same thing can be happen in the rod model. So, in this sense, the discontinuous transition found for k-mers in the mean-field solution on the triangular Husimi lattice is consistent with the 3-state Potts scenario for the triangular lattice. We notice, for instance, that a similar scenario was recently found for an associating lattice gas model, which exhibits a continuous transition in the 3-state Potts class between a fluid and a high-density liquid phase, as revealed by Monte Carlo simulations and transfer matrix calculations \cite{furlanALG,ibagon21}, but it was found to be discontinuous in Husimi lattice solutions \cite{tiago10,furlanALG}. Furthermore, we recall that numerical simulations of the rod model on the cubic lattice \cite{vdr17} suggested the isotropic-nematic transition to be continuous for $k \ge 7$, but the possibility of a weak discontinuous first-order transition, as is found for the 3-state Potts model on this lattice \cite{w82}, was not ruled out, although it is not clearly supported by the simulational results. Once again, the isotropic-nematic transitions found in the early studies via mean-field approximations of the three-dimensional models mentioned before \cite{o49,f56,z63} were also discontinuous.

If we consider the critical densities at which the transition happen in our calculations for the square case, they are always between the ones found on the Bethe lattice and the estimates coming from the simulations. The same is true for the entropies in the full lattice limit, so that for these parameters the Husimi lattice results are closer to the ones on the corresponding Bravais lattices than the ones coming from the solution on the Bethe lattice, as expected. On the other hand, the coexistence densities obtained in the triangular case display an opposite trend, suggesting that they converge non-monotonically to the triangular lattice results as the mean-field approximation is improved. Another point where there was no improvement in the Husimi lattice solutions when compared to the ones on the Bethe lattice was the minimum value of $k_{min}$ for which the phase transition was observed, which is equal to four in all cases, as opposed to $k_{min}=7$ for the square and triangular lattices. Calculations in the full lattice limit \cite{rso22} on hierarchical lattices built with larger square cells, such as the Kobayashi-Susuki lattice \cite{ks93}, indicate that larger values of $k_{min}$ are found when the size of the cells grows. Therefore, it is interesting to extend this study to include vacancies in the configurations of the model on such generalized Husimi lattices to find out the complete thermodynamic behavior of rigid rods on them. In a similar vein, investigating the $k$-mers on such lattices built with triangular lattice cells can be also very important to verify whether the isotropic-nematic transition turns out to be continuous at some point, as well as to estimate the entropies in the full-packing limit, since it seems that they have never been calculated in the literature for the triangular lattice, beyond the dimer case. We are currently working in these projects.

\section*{ACKNOWLEDGMENTS}

This work used computational resources of the ``Centro Nacional de Processamento de Alto Desempenho" in São Paulo (CENAPAD-SP, FAPESP). 
The authors acknowledge financial support by FAPERJ, FAPEMIG, CAPES and CNPq (Brazilian agencies).

\appendix

\section{Solution of the model on the Husimi lattice built with triangles}
\label{secHLtri}

\begin{figure}[t]
\centering
\includegraphics[width=8.50cm]{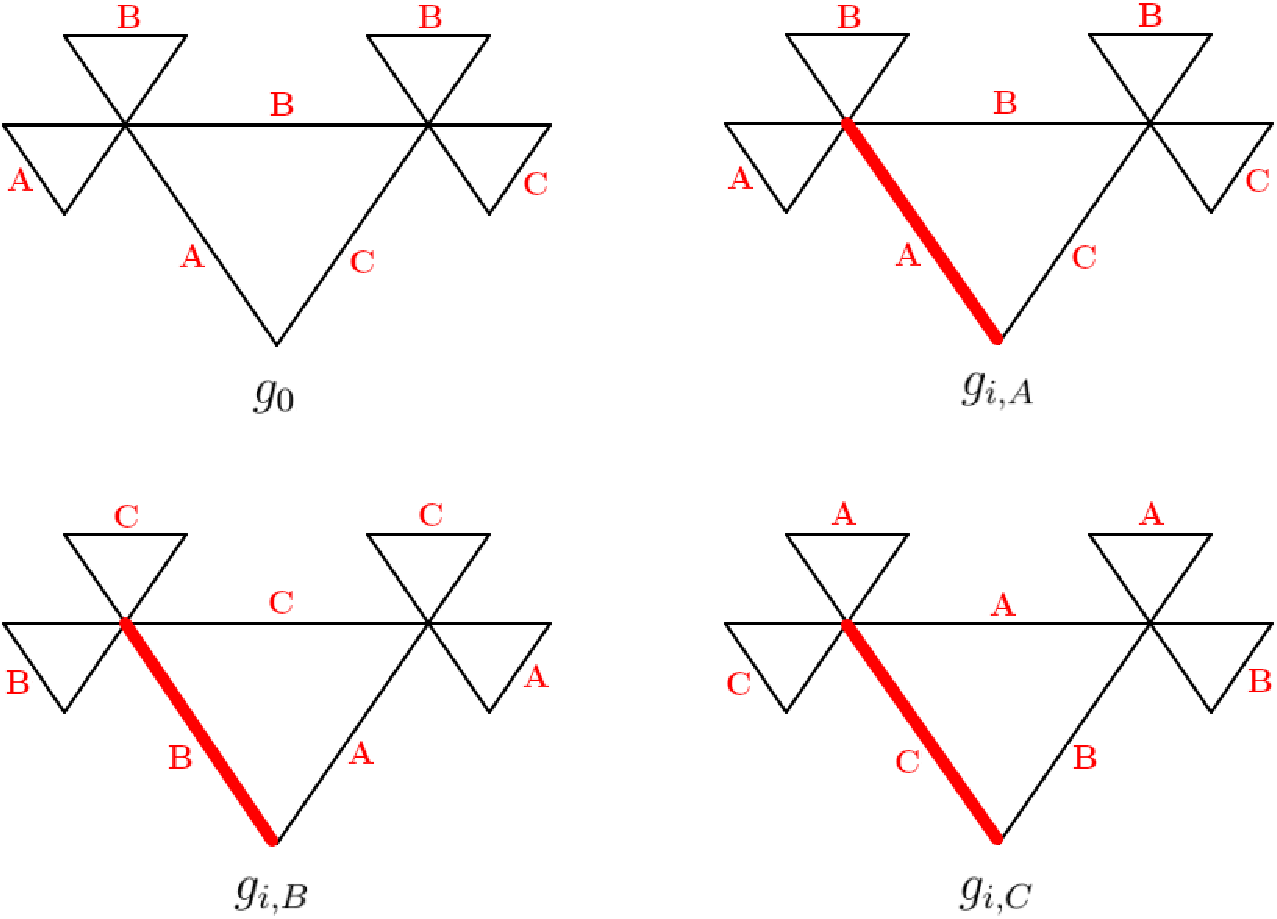}
\caption{Possible states of the root sites for rods placed on the triangular HL. The thicker (red) lines represent a reaching rod with $i$ monomers incorporated to it. The lattice directions ($A$, $B$ and $C$) are also indicated.}
\label{FigstaHLtri}
\end{figure}

We may obtain the solution of the rod problem on the triangular HL [as defined in Fig. \ref{FigHT}(b)] following the same lines as in the square HL case discussed in Sec. \ref{SecHL}. The main differences are that now we have three lattice directions --- which will be labeled as $A$, $B$ and $C$ here --- and two sub-cactuses are attached to each of the two non-rooted sites of a rooted triangle, in order to obtain a lattice with coordination $q=6$; see Figs. \ref{FigHT}(b) and \ref{FigstaHLtri}. Since there are three lattice directions, but only two edges arriving at a root site, at first, one should distinguish among the situations where such edges are of type $AB$, $BC$ or $CA$. To keep the notation as simple as possible, we will name the ppf's according to the label of the direction opposite (i.e., not connecting) to the root site. Thereby, for the configurations with no reaching rod (from above) in the root site, we will have the ppf's $a_0$, $b_0$ and $c_0$. In the same way, we shall have the ppf's $b_{(i,A)}$ and $c_{(i,A)}$; $a_{(i,B)}$ and $c_{(i,B)}$; and $a_{(i,C)}$ and $b_{(i,C)}$, when the reaching rod has $i$ monomers incorporated to it in direction $A$, $B$ or $C$. By defining the sums:
\begin{subequations}
\begin{eqnarray}
  s_A = b_0 c_0 + z_A \sum_{n=0}^{k-1}b_{(n,A)}c_{(k-1-n,A)} + z_B b_0 c_{(k-1,B)} + z_C c_0 b_{(k-1,C)}, \\
  s_B = a_0 c_0 + z_A a_0 c_{(k-1,A)} + z_B \sum_{n=0}^{k-1}a_{(n,B)}c_{(k-1-n,B)} + z_C c_0 a_{(k-1,C)}, \\
  s_C = a_0 b_0 + z_A a_0 b_{(k-1,A)} + z_B b_0 a_{(k-1,B)} + z_C \sum_{n=0}^{k-1}  a_{(k-1-n,C)} b_{(n,C)},
\end{eqnarray}
\label{eqRRsRHtriSG}
\end{subequations}
the recursion relations (RRs) for the ppf's related to the opposite direction $A$ are given by
\begin{subequations}
\begin{eqnarray}
  a'_0&=&s_B s_C + z_A^2 a_0^2 \sum_{n=0}^{k-2}b_{(n,A)}c_{(k-2-n,A)}, \\
  a'_{(i,B)}&=&z_B b_0 a_{(i-1,B)} s_B,\\
  a'_{(i,C)}&=&z_C c_0 a_{(i-1,C)} s_C.
\end{eqnarray}
\label{eqRRsRHtri}
\end{subequations}
where $i=1,\ldots,k-1$, for $k \geqslant 2$. The analogous RRs for $b$ and $c$ can be obtained from these expressions through cyclic permutations of type $a \rightarrow b$, $b \rightarrow c$ and $c \rightarrow a$, along with $A \rightarrow B$, $B \rightarrow C$ and $C \rightarrow A$.

Then, by defining the ratios $F_{(i,B)}=a_{(i,B)}/a_0$, $F_{(i,C)}=a_{(i,C)}/a_0$, $G_{(i,A)}=b_{(i,A)}/b_0$, $G_{(i,C)}=b_{(i,C)}/b_0$, $H_{(i,A)}=c_{(i,A)}/c_0$ and $H_{(i,B)}=c_{(i,B)}/c_0$, one readily obtains a set of $(6k-6)$ RRs for them, whose fixed points define the phases of the system. It turns out, however, that such RRs always converge to fixed points where $G_{(i,A)}=H_{(i,A)}$, $F_{(i,B)}=H_{(i,B)}$ and $F_{(i,C)}=G_{(i,C)}$, demonstrating that there is no need to distinguish between the different directions of the opposite edges. Therefore, we can simplify the solution, by considering a single ppf $g_0$ (instead of $a_0$, $b_0$ and $c_0$) when no rod arrives at the root site (from above), and ppf's $g_{(i,j)}$ if a rod with $i=1,\ldots,k-1$ monomers incorporated reaches the root site in direction $j=A,B$ or $C$. See Fig. \ref{FigstaHLtri} for a illustration of these definitions. In this way, the number of ppf's reduces to only $3k-2$, with the related RRs given by
\begin{subequations}
\begin{eqnarray}
  g'_0&=&s_A s_C + z_B^2 g_0^2 \sum_{n=0}^{k-2}g_{(n,B)}g_{(k-n-2,B)}, \\
  g'_{(i,A)}&=&z_A g_0 g_{(i-1,A)} s_A,\\
  g'_{(i,B)}&=&z_B g_0 g_{(i-1,B)} s_B,\\
  g'_{(i,C)}&=&z_C g_0 g_{(i-1,C)} s_C,
\end{eqnarray}
\label{eqRRsRHtri1}
\end{subequations}
where the sums defined in Eqs. \ref{eqRRsRHtriSG} simplify to
\begin{subequations}
\begin{eqnarray}
  s_A = g_0^2 + z_A \sum_{n=0}^{k-1}g_{(n,A)}g_{(k-n-1,A)} + z_B g_0 g_{(k-1,B)} + z_C g_0 g_{(k-1,C)}, \\
  s_B = g_0^2 + z_A g_0 g_{(k-1,A)} + z_B \sum_{n=0}^{k-1}g_{(n,B)}g_{(k-n-1,B)} + z_C g_0 g_{(k-1,C)}, \\
  s_C = g_0^2 + z_A g_0 g_{(k-1,A)} + z_B g_0 g_{(k-1,B)} + z_C \sum_{n=0}^{k-1}g_{(n,C)}g_{(k-n-1,C)}.
\end{eqnarray}
\label{eqRRsRHtriS}
\end{subequations}
Note that $g_{(0,A)}=g_{(0,B)}=g_{(0,C)}=g_{0}$ in these equations. Then, we may defined the (new) ratios of ppf's as:
\begin{equation}
    A_i = \frac{g_{(i,A)}}{g_0}, \quad B_i = \frac{g_{(i,B)}}{g_0}, \quad  \text{and} \quad C_i = \frac{g_{(i,C)}}{g_0},
\end{equation}
so that equations \ref{eqRRsRHtri1} yield a set of $3(k-1)$ RRs for them. For $k \leq 3$, and $z_A=z_B=z_C=z$, such RRs always converge to a fixed point of type $A_i=B_i=C_i$, corresponding to the isotropic phase. For $k \geq 4$, besides the isotropic fixed point, a nematic one is stable for large $z$, where, e.g., $A_i \neq B_i = C_i$.

Considering the simplified solution, the partition function of the system, obtained by attaching six sub-cactuses to the three vertices of the central triangle, reads:
\begin{eqnarray}
 Y = s_A s_B s_C &+& z_A^2 g_0^2 s_A \sum_{n=0}^{k-2}g_{(n,A)}g_{(k-2-n,A)} + z_B^2 g_0^2 s_B \sum_{n=0}^{k-2}g_{(n,B)}g_{(k-2-n,B)} \nonumber \\ 
 &+& z_C^2 g_0^2 s_C \sum_{n=0}^{k-2}g_{(n,C)}g_{(k-2-n,C)} = g_0^6 y,
\end{eqnarray}
where $y$ only depends on $z$ and on the ratios $A_i$, $B_i$ and $C_i$ at their fixed points, in the thermodynamic limit.

The density of monomers per site (at the vertices of the central triangle) belonging to rods in direction $j=A,B$ or $C$ is given by:
\begin{equation}
 \rho_j = \frac{z_j}{3y} \frac{\partial y}{\partial z_j},
\end{equation}
and, then, the total density is $\rho = \rho_A+\rho_B+\rho_C$. Following Ref. \cite{drs11}, the order parameter for the isotropic-nematic transition will be defined as
\begin{equation}
 \Psi = \frac{\rho_A - \rho_B}{\rho},
\end{equation}
assuming that $\rho_B=\rho_C$ always.

As demonstrated, e.g., in Ref. \cite{furlanALG}, the bulk free energy per site, $\phi_b$, for the triangular HL is
\begin{equation}
\phi_b = -\frac{k_B T}{3} \ln\left[\frac{Y'}{Y^4} \right] = -k_B T \ln\left[\frac{f_0^2}{y} \right],
\end{equation}
so that
\begin{equation}
\varphi_b = \frac{\phi_b}{k_B T} = - \ln\left[\frac{f_0^2}{y} \right],
\end{equation}
with $f_0$ being the convergent part of $g'_0$, once we write $g'_0=g_0^4 f_0$.


\begin{thebibliography}{99}
\bibitem{cmc05}N. Clisby and B. M. McCoy, Pramana {\bf 64}, 775 (2005).
\bibitem{ps88}P. A. Pearce and K. A. Seaton, J. Stat. Phys. {\bf 53}, 1061 (1988).
\bibitem{fal07}H. C. M. Fernandes, J. J. Arenzon, and Y. Levin, J. Chem. Phys. {\bf 126}, 114508 (2007).
\bibitem{b80}R. J. Baxter, J. Phys. A: Math. Gen. {\bf 13}, L61 (1980).
\bibitem{vn99}A. Verberkmoes and B. Nienhuis, Phys. Rev. Lett. {\bf 83}, 3986 (1999).
\bibitem{bsg09}B. C. Barnes, D. W. Siderius, and L. D. Gelb, Langmuir {\bf 25}, 6702 (2009).
\bibitem{m05}Y. Maeda, T. Niori, J. Yamamato, and H. Yokoyoma, Thermochim. Acta {\bf 431}, 87 (2005).

\bibitem{binary} M. Dijkstra, Phys. Rev. E {\bf 58}, 7523 (1998); M. Schmidt, J. Phys.: Condens. Matter {\bf 16}, L351 (2004); J. M. Brader and R. L. C. Vink, J. Phys.: Condens. Matter {\bf 19}, 036101 (2007); D. Frenkel and A. A. Louis, Phys. Rev. Lett. {\bf 68}, 3363 (1992); R. Dickman and G. Stell, J. Chem. Phys. {\bf 102}, 8674 (1995); R. van Roij, B. Mulder, and M. Dijkstra, Physica A {\bf 261}, 374 (1998); H. H. Wensink, G. J. Vroege, and H. N. W. Lekkerkerker, J. Chem. Phys. {\bf 115}, 7319 (2001); S. Dubois and A. Perera, J. Chem. Phys. {\bf 116}, 6354 (2002); S. Varga, A. Galindo, and G. Jackson, J. Chem. Phys. {\bf 117}, 7207 (2002); M. Schmidt and A. R. Denton, Phys. Rev. E {\bf 65}, 021508 (2002); Y. Mart\'inez-Rat\'on, E. Velasco, and L. Mederos, Phys. Rev. E {\bf 72}, 031703 (2005); D. de las Heras, Y. Mart\'inez-Rat\'on, and E. Velasco, Phys. Rev. E {\bf 76}, 031704 (2007); N. T. Rodrigues and T. J. Oliveira, J. Chem. Phys. {\bf 151}, 024504 (2019); N. T. Rodrigues and T. J. Oliveira, Phys. Rev. E {\bf 100}, 032112 (2019).

\bibitem{ternary} H. M. Schaink, Physica A {\bf 210}, 113 (1994); S. B. Yuste, A. Santos, and M. L\'opez de Haro, J. Chem. Phys. {\bf 108}, 3683 (1998); A. Santos, S. B. Yuste, and M. L\'opez de Haro, J. Chem. Phys. {\bf 117}, 5785 (2002); A. Malijevsk\'y, A. Malijevsk\'y, S. B. Yuste, A. Santos, and M. L\'opez de Haro, Phys. Rev. E {\bf 66}, 061203 (2002); A. Malijevsk\'y, S. Lab\'ik, and A. Malijevsk\'y, Phys. Chem. Chem. Phys. {\bf 6}, 1742 (2004); C. N. Patra and S. K. Ghosh, J. Chem. Phys. {\bf 118}, 3668 (2003); Y.-X. Yu, J. Chem. Phys. {\bf 121}, 1535 (2004); N. T. Rodrigues and T. J. Oliveira, Phys. Rev. E {\bf 101}, 062102 (2020).

\bibitem{vl92}G. J. Vroege and H. N. W. Lekkerkerker, Rep. Prog. Phys. {\bf 55}, 1241 (1992).

\bibitem{o49}L. Onsager, Ann. N.Y. Acad. Sci. {\bf 51}, 627 (1949).
\bibitem{f56}P. J. Flory, Proc. R. Soc. 234, {\bf 60} (1956).
\bibitem{z63}R. Zwanzig, J. Chem. Phys. {\bf 39}, 1714 (1963).
\bibitem{k61}P. W. Kasteleyn, Physica {\bf 27}, 1209 (1961); J. Math. Phys. {\bf 4}, 287 (1963).
\bibitem{tf61}H. N. V. Temperley and M. E. Fisher, Phil. Mag. {\bf 6}, 1061 (1961); M. E. Fisher, Phys. Rev. {\bf124}, 1664 (1961).
\bibitem{l67}E. H. Lieb, J. Math. Phys. {\bf 8}, 2339 (1967).
\bibitem{wp21}N. Wilkins and S. Powell, Phys. Rev. E {bf 104}, 014145 (2021).

\bibitem{Nagle66} J. F. Nagle, Phys. Rev. {\bf 152}, 190 (1966).
\bibitem{Phares85} A. J. Phares and F. J. Wunderlich, J. Math. Phys. {\bf 27}, 1099 (1985).
\bibitem{Kenyon00} R. Kenyon, CRM Monogr. Ser. Amer. Math. Soc. {\bf 13}, 307 (2000).
\bibitem{Fendley02} P. Fendley, R. Moessner, and S. L. Sondhi, Phys. Rev. B {\bf 66}, 214513 (2002).

\bibitem{Wu06} F. Y. Wu, Int. Jour. Mod. Phys. B {\bf 20}, 5357 (2006).
\bibitem{dc08} D. Dhar and S. Chandra, Phys. Rev. Lett. {\bf 100}, 120602 (2008).

\bibitem{gdj07}A. Ghosh, D. Dhar, and J. L. Jacobsen, Phys. Rev. E {\bf 75}, 011115 (2007).
\bibitem{rsd23}L. R. Rodrigues, J. F. Stilck, and W. G. Dantas, Phys. Rev. E {\bf 107}, 014115, (2023). 
\bibitem{p21}P. M. Pasinetti, A. J. Ramirez-Pastor, E. E. Vogel, and G. Saravia, Phys. Rev. E {\bf 104}, 054136 (2021).
\bibitem{rso22}N. T. Rodrigues, J. F. Stilck, and T. J. Oliveira, Phys. Rev. E {\bf 105}, 024132 (2022).

\bibitem{gd07}A. Ghosh and D. Dhar, Europhys. Lett. {\bf 78}, 20003 (2007).

\bibitem{mf08}D. A. Matoz-Fernandez, D. H. Linares, and A. J. Ramirez-Pastor, Europhys. Lett. {\bf 82}, 50007 (2008).
\bibitem{mf08p}D. A. Matoz-Fernandez, D. H. Linares, and A. J. Ramirez-Pastor, J. Chem. Phys. {\bf 128}, 214902 (2008).
\bibitem{fv09}T. Fischer and R. L. C. Vink, Europhys. Lett. {\bf 85}, 56003 (2009).
\bibitem{mf08pp}D. Matoz-Fernandez, D. Linares, and A. Ramirez-Pastor, Physica A {\bf 387}, 6513 (2008).

\bibitem{ds13}M. Disertori and A. Giuliani, Commun. Math. Phys. {\bf 323}, 143 (2013).

\bibitem{k12}J. Kundu, R. Rajesh, D. Dhar, and J. F. Stilck, AIP Conf. Proc. {\bf 1447}, 113 (2012).
\bibitem{k13}J. Kundu, R. Rajesh, D. Dhar, and J. F. Stilck, Phys. Rev. E {\bf 87}, 032103 (2013).
\bibitem{sdr22}A. Shah, D. Dhar, and R. Rajesh, Phys. Rev. E {\bf 105}, 034103 (2022). 
\bibitem{drs11}D. Dhar, R. Rajesh, and J. F. Stilck, Phys. Rev. E {\bf 84}, 011140 (2011).
\bibitem{b82}R. J. Baxter, \textit{Exactly Solved Models in Statistical Mechanics} (Academic Press, London, 1982).
\bibitem{h50}K. Husimi, J. Chem. Phys. {\bf 18}, 682 (1950). 
\bibitem{t76}T. Tsuchyia, Prog. Theor. Phys. {\bf 56}, 741 (1976).
\bibitem{dr21}D. Dhar and R. Rajesh, Phys. Rev. E {\bf 103}, 042130 (2021).


\bibitem{Gujrati} P. D. Gujrati, Phys. Rev. Lett. {\bf 74}, 809 (1995).
\bibitem{MinosJurgen}M. A. Neto and J. F. Stilck, J. Chem. Phys. \textbf{138}, 044902 (2013).
\bibitem{tiagoPol} T. J. Oliveira, J. Phys. A: Math. Theor. \textbf{49}, 155001 (2016).
\bibitem{furlanALG} A. P. Furlan, T. J. Oliveira, J. F. Stilck, and R. Dickman, Phys. Rev. E \textbf{100}, 022109 (2019).


\bibitem{ro21} N. T. Rodrigues and T. J. Oliveira, Phys. Rev. E \textbf{103}, 032153 (2021).

\bibitem{vdr17} N. Vigneshwar1, D. Dhar, and R. Rajesh, J. Stat. Mech., 113304 (2017).
\bibitem{w82} F. Y. Wu, Rev. Mod. Phys. \textbf{54}, 235 (1982).
\bibitem{ks93}H. Kobayashi and M. Suzuki, Physica A \textbf{199}, 619, (1993).

\bibitem{ibagon21} I. Ibagon, A. P. Furlan, T. J. Oliveira, and R. Dickman, Phys. Rev. E \textbf{104}, 064120 (2021).
\bibitem{tiago10} T. J. Oliveira, J. F. Stilck, and M. A. A. Barbosa, Phys. Rev. E \textbf{82}, 051131 (2010).
\bibitem{ostilli} M. Ostilli, Phys. A \textbf{391}, 3417 (2012).
\end{thebibliography}
\end{document}